\documentclass[a4paper,10pt]{article}

\usepackage{amsfonts,amsmath,graphicx}
\usepackage{comment}
\usepackage{wrapfig}
\usepackage{epsfig}

\def\d{{\rm d}}

\usepackage{multicol}[2]

\title{Study of some parameters interstellar transport using of magnetic umbrella}

\author{Martin \v{C}ERM\'{A}K \bigskip
\\
\normalsize\it Department of Physical Electronics, Faculty of Science, Masaryk University\\
\normalsize\it Brno, Kotl\'{a}\v{r}sk\'{a} 2, Czech Republic\\
\normalsize\it 63855@mail.muni.cz 
}

\begin{document}
\maketitle

\begin{abstract}
Interstellar transport is an object of interest in many sci-fi stories. In history a lot of sci-fi predictions have turned into reality, such as communications satellites, deep-sea submarines and journies to the moon. In this work we study some physical parameters of a space ship which uses a magnetic umbrella. Our spaceship generates a magnetic field in its neighborhood and captures charged protons into a magnetic trap. These particles are taken into a fusion reactor. The obtained energy and waste in form of helium are used as a fuel in an ion engine. With the help of elementary physics we can work out the basic physical parameters of the ship, e.g. maximal velocity, acceleration of the ship or acceleration time period.
\end{abstract}





\section{Introduction}
Interstellar travel has many unsolved difficulties causing man's dream to explore new undiscovered worlds to be far in the future. Problems like sources of energy, a shield against micro meteors and engines with sufficient and long term traction still haven't been solved yet. The interstellar ramjet is trying to solve some of these problems.

There are a many advantages in using a interstellar ramjet in space travel. The main one is that the spacecraft doesn't have to carry its own fuel. This can reduce the weight of the spaceship, the requirements regarding traction of the engines and therefore initial acceleration can be much higher. The interstellar ramjet was for the first time studied by Bussard \cite{bus}. The next utilization of this system can be as a solar-sail \cite{angeo-25-755-2007} or space parachute. The principle of a solar-sail can be used to gain initial speed, which is needed for the interstellar ramjet to work. On the other hand, when the spaceship arrives at an alien star system, the magnetic umbrella can be used as a parachute which decreases entry velocity.

Previous studies considered standard type of interstellar ramjet spaceship \cite{2007AcAau..61..817S}. Ramjet spaceship does not stop particles (due to spaceship frame). Then particles have non-zero average velocity while they fly through nuclear fusion reactor. In this work we consider different type of the propulsion process with lower speed limitation for the spaceship. On the other hand, our device could be more real to construct then typical ramjet spaceship.

\section{Technical principle of magnetic umbrella}
The principle of the magnetic umbrella is similar to how a magnetic field around the earth captures particles of the solar wind. There are technical problems how to make a magnetic field with a long range. The problem can be solved by new materials with extra properties as for example superconductors, which have been developed and studied in the last few decades. These super-materials with low mass could use low energy to generate a strong magnetic field. The magnetic field will catch charged particles moving in the proximity of the umbrella and transport them into the ship. The aurora is made in a similar way. At the moment of being caught, the particles emit a photon which takes away part of the kinetic energy.

Interstellar space is full of neutral and charged particles. Most of them are atoms or ions of Hydrogen and electrons.
The density of protons (positive ions of the simplest hydrogen) in interstellar space is around 
$\eta_{p}=0.04-0.07\; \rm{cm}^{-3}$
 \cite{Izmodenov:1998ws}.

The number of captured protons grows with the velocity of the ship. On the other side particles apply a resistant force on the ship.
Captured particles are taken into a fusion reactor where the binding energy is transformed into kinetic energy of the particles and the ship. Waste products from this reaction are nuclei of the Helium. These nuclei are taken into the engines and leave the spacecraft at a high velocity.

\section{Maximal velocity of the spacecraft}
Previous studies considered speed limit of interstellar ramjet spaceship \cite{2007AcAau..61..817S} given by thermal lost and lost in the interstellar medium. Maximal velocity with zero lost is speed of light. In this work we consider different type of the propulsion process with speed limitation for the spaceship given by resistant force.

Consider that a spacecraft with mass $M$ moves due to interstellar gas with a velocity $v.$ 
In our calculations we can choose a frame of reference connected with the spacecraft. In this case, interstellar gas moves with a velocity $v$ due to the ship.

Let's assume that one of the particles with a rest mass $m_{0}$ of ionized gas is captured and emits a photon. The laws of conservation of momentum and energy say that the momentum and energy of the system has to be the same before and after the reaction. For accuracy assume that the particle is moving with a relativistic velocity. Kinetic energy and momentum of the particle can be expressed as
$$
E_{kp}=m_{0}c^{2}\left( \gamma-1\right), \;\; p_{p}=\gamma m_{0}v, \;\; 1/\gamma=\sqrt{1-\frac{v^{2}}{c^{2}}}.
$$
Before capture of the particle, the spaceship is at rest; its kinetic energy is equal to zero.

Let's assume that the mass of the spaceship is much larger than the mass of the particle. For this reason the kinetic energy and momentum of the ship with the particle after the capture are nonrelativistic
$$
E_{k p+s}=\frac{1}{2}\left(m_{0}+M\right)V_{1}^{2},\;\; p_{p+s}=\left(m_{0}+M\right)V_{1},
$$
where $V_{1}$ is velocity of the ship (with particle) after capture.
In our calculations energy $E_{f}$ and momentum $p_{f}=\frac{E_{f}}{c}$ of the photon is also important.

As we mentioned above energy before and after has to be same
\begin{eqnarray}
\def\arraystretch{1.5}\begin{array}{l}
m_{0}c^{2}\left( \gamma-1\right)=\frac{1}{2}\left(m_{0}+M\right)V_{1}^{2}+E_{f},\\
\gamma m_{0}v=\left(m_{0}+M\right)V_{1}+\frac{E_{f}}{c}.
\end{array}
\label{1}
\end{eqnarray}
In fact the first equation is a scalar equation and the second is a vector equation. Consider that the terms in the second equation $v$, $V_{1},$ $\frac{E_{f}}{c}$ are only positive numbers (for simplicity we will use absolute values of $\left\vert v\right\vert$).

Note that the second equation is only approximation. In real case emission of the photon will not be only in the direction of the ship speed. Moreover during the process of capturing will be emitted more than one photon. The photon radiation will be strongly dependent on magnetic field and velocity of the captured proton. For more precise calculations can be term $\frac{E_{f}}{c}$ in the second equation replaced by $A\frac{E_{f}}{c}$, where $A\in[0;1]$. In this work we consider only ideal case $A=1,$ where radiation is emitted only in the direction of the ship speed.

In the above-mentioned equations there are two unknowns $V_{1}$ and $E_{f}.$ It is easy to express the velocity of the ship $V_{1}$
$$ 
V_{1}=c\left[1-\sqrt{1+\frac{2m_{0}}{m_{0}+M}\left( \sqrt{\frac{1-\frac{\left\vert v\right\vert}{c}}{1+\frac{\left\vert v\right\vert}{c}}}-1\right) } \right].
$$

As we note, this solution is only for $v>0$ and $V_{1}>0$. For negative values of $v$ this solution must be multiplied by $-1,$ so that $V_{1}(v)=-V_{1}(-v).$

For small term $a$ in the root we can use Taylor expansion:$\sqrt{1+2a}\approx 1+a,$ in this spirit we can say that $\frac{2m_{0}}{m_{0}+M}\approx\frac{2m_{0}}{M}$ is very small, because $M\gg m_{0}$. Therefore we can write
\begin{eqnarray}
V_{1}\approx c\frac{m_{0}}{M}\left( 1-\sqrt{\frac{1-\frac{\left\vert v\right\vert}{c}}{1+\frac{\left\vert v\right\vert}{c}}}\right).\label{2}
\end{eqnarray}
This is the velocity of the ship after the capture of a proton.

\medskip
The maximal velocity is reached if the velocity which is lost after capturing the proton $V_{1}$ is equal to the velocity which is obtained from the engines $V_{2}$. From this reasoning we can assume that the velocity of the rocket after the reaction is again equal to zero. That part of the particle's rest energy is transformed into kinetic energy in the fusion reactor. Therefore in this case we will include rest energy of the particle before $m_{0}$ and after $m'_{0}$ the reaction. The difference between $\left(m_{0}-m'_{0}\right)c^{2}$ is the energy obtained from nuclear fusion.

The energy and momentum of the rockets before nuclear fusion is
$$
E_{p+s}=\frac{1}{2}\left(m_{0}+M\right)V_{2}^{2}+m_{0}c^{2}, p_{p+s}=\left(m_{0}+M\right)V_{2}.
$$
The energy and momentum after the reaction is
$$
E_{p}=\gamma' m'_{0}c^{2}, \;\; p_{p}=\gamma' m'_{0}v', \;\; 1/\gamma'=\sqrt{1-\frac{v'^{2}}{c^{2}}}.
$$
If the effectiveness of the process is not $100$ percent, part of the energy is transformed into thermal energy $E_{t}.$

The laws of conservation of momentum and energy can be written in the form
\begin{eqnarray}
\def\arraystretch{1.5}\begin{array}{l}
\frac{1}{2}\left(m_{0}+M\right)V_{2}^{2}+m_{0}c^{2}=\gamma' m'_{0}c^{2}+E_{t},\\
\left(m_{0}+M\right)V_{2}=\gamma'm'_{0}v'.
\end{array}\label{3}
\end{eqnarray}
From the second equation the velocity of the particle after the reaction $v'$ can be expressed as
\begin{eqnarray}
v'=\frac{\frac{m_{0}+M}{m'_{0}}V_{2}}{\sqrt{1+\left(\frac{ m_{0}+M}{m'_{0}}\right)^{2}\frac{V_{2}^{2}}{c^{2}}}}.\label{v}
\end{eqnarray}
This can be substituted into the first equation (\ref{1})
$$
\frac{m_{0}+M}{2}V_{2}^{2}+m_{0}c^{2}=m'_{0}c^{2}\sqrt{1+\left(\frac{ m_{0}+M}{m'_{0}}\right)^{2}\frac{V_{2}^{2}}{c^{2}}}+E_{t}.
$$
And finally velocity $V_{2}$ is
$$
V_{2}^{2}=\frac{2\left[E_{t}+Mc^{2}-\sqrt{\left( E_{t}+Mc^{2}\right)^{2}-\left(m_{0}c^{2}-E_{t}\right)^{2}+\left( m'_{0}c^{2}\right)^{2} } \right]}{m_{0}+M}.
$$
It can be expected that $\left( E_{t}+Mc^{2}\right)^{2}\gg\left(m_{0}c^{2}-E_{t}\right)^{2}-\left( m'_{0}c^{2}\right)^{2},$ $Mc^{2}\gg E_{t},$ and $M\gg m_{0},$ therefore we can use Taylor expansion again
\begin{equation}
V_{2}^{2}\approx\frac{\left(m_{0}c^{2}-E_{t}\right)^{2}-\left( m'_{0}c^{2}\right)^{2}}{M^{2}c^{2}}.
\label{4}
\end{equation}
The same solution can be obtained if the energy which is released corresponds to all the rest energy of the particle ($m'_{0}c^{2}=0$) in the process of annihilation. 

If the spacecraft reaches maximal speed, the change of velocity must be zero $V_{2}-V_{1}=0.$
A combination of (\ref{2}) and (\ref{4}) leads to the equation of maximal velocity.
\begin{equation}
\left( 1-\sqrt{\frac{1-\frac{\left\vert v\right\vert}{c}}{1+\frac{\left\vert v\right\vert}{c}}}\right)^2 \approx\frac{\left(m_{0}c^{2}-E_{t}\right)^{2}-\left( m'_{0}c^{2}\right)^{2}}{\left( m_{0}c^{2}\right) ^{2}}\equiv\alpha^{2},
\label{5}
\end{equation}
then
$$
\frac{\left\vert v\right\vert}{c}\approx\frac{1-\left(1-\alpha\right)^{2}}{1+\left(1-\alpha\right)^{2}}.
$$

As an example we will assume a proton-proton fusion reaction. A nuclear reaction takes the form 
$4p+2e^{-}\rightarrow ^{4}\hspace{-2mm} He+2\nu_{e}+26.73\rm{MeV}$ \cite{Brass}. The rest mass of the proton is $938\rm{MeV}$. The energy of the neutrinos and the remaining mass of the electron ($0.51\rm{MeV}$) are negligible due to the energy of the reaction. We can say that during a nuclear reaction  $0,7\%$ energy of the proton is released, so then $m'_{0}=0.993~m_{0}$. Let us assume that heat losses and other energy losses are negligible $E_{t}=0$. Maximal velocity of the spaceship expressed from (\ref{5}) is $v=0.125c.$

As we mentioned above maximal velocity of the spaceship is given by (\ref{5}). Equation (\ref{5}) will change for annihilation of antiprotons and protons into the form
$$
\frac{\left\vert v\right\vert}{c}=\frac{\left( m_{0}c^{2}\right)^{2}-E_{t}^{2}}{\left( m_{0}c^{2}\right)^{2}+E_{t}^{2}}.
$$
It is obvious that maximal velocity is limited only by the efficiency of the process. Similar results was found in \cite{2007AcAau..61..817S}.

As well as atoms and ions, interstellar matter contains a small number of high energy antiprotons and positrons. These particles are components of cosmic rays. Antiproton$/$proton ratio is a function of the energy of the particles. For higher energies this ratio increases. For kinetic energies $1-10\rm{GeV}$ this ratio is about $10^{-6}-2\cdot 10^{-4}$ \cite{Adriani:2010rc}. Maximal antiproton flux is about $0.025(\rm{GeV m}^{2} \rm{s}\;  \rm{sr})^{-1}$ for kinetic energies $2-3\rm{GeV}$ \cite{Moskalenko:2001ya}.

These numbers are too small for effective propulsion of the spaceship. For an exact calculation we should know the function of flux as a function of energy.

\section{Acceleration of the spacecraft}
Relativistic acceleration is defined \cite{Landau}
$$
a=\frac{\d^{2} x}{\d \tau^{2}}=\frac{\d}{\d \tau}\left(\frac{\d x}{\d t}\frac{\d t}{\d \tau}\right)=\frac{\d}{\d \tau}\left(\gamma v\right)=\frac{\d v}{\d \tau}\gamma^{3}, v=\frac{\d x}{\d t},
$$
where $x$ is the spatial coordinate. Dependence between coordinate time $t$ and proper time $\tau$ is 
$$
\d\tau=\frac{1}{\gamma}\d t.
$$

Every particle will change the velocity of the ship. The change in velocity is $\Delta V=V_{2}-V_{1},$ where $V_{2}$ and $V_{1}$ are given by (\ref{2}) and (\ref{4}). The change of the velocity is a function of the actual velocity $\Delta V(v)$.

The number of particles that the spacecraft can catch $N$ is dependent on actual velocity $v,$ concentration of the protons $\eta_{p}$ and the area of the magnetic umbrella $S.$ If the ship is moving at a velocity near to the speed of light, concentration of the particles grows with $\gamma.$ Number of particles $\d N$ which meet the magnetic field in some small period of proper time $\d \tau$ is
$$
\d N=S\d x\eta_{p}=\eta_{p}vS\d t=\eta_{p}\gamma vS\d \tau. 
$$

A small change in the velocity is given by changing the velocity for one particle multiplied by the number of particles in a period of proper time
\begin{equation}
\d v=dN\Delta V=\Delta V\eta_{p}\gamma vS\d \tau.
\label{6x}
\end{equation} 
By definition, acceleration is
\begin{equation}
a=\Delta VSv\eta_{p}\gamma^{4}.
\label{6}
\end{equation}
This is the acceleration which an observer in the ship would feel.

In the case of a solar-sail or space parachute the spacecraft can use the velocity of the particles for higher deceleration. In this case the change of velocity $\Delta V$ in the equation (\ref{6}) is given by $\Delta V=V_{2}+\left\vert V_{1}\right\vert$, because $V_{1}$ is oriented in the same direction as $V_{2}$.
\bigskip

Functions $a(v)$ and $\gamma(v)$ are functions of the velocity of the ship due to the interstellar particles (or particles due to the ship). Time period of acceleration (proper time measured on board the ship) is dependent on the velocity $v$ and can easily be calculated as 
\begin{equation}
\Delta \tau=\int_{v_{0}}^{v_{fin}}\frac{\gamma^{3}(v)\d v}{a(v)}=\int_{v_{0}}^{v_{fin}}\frac{\d v}{\gamma(v)\Delta V(v)Sv\eta_{p}}.\label{6x}
\end{equation}

Consider three different examples. 
\begin{itemize}
\item
In the first case, the ship will use a combination of engines with the effect of a solar-sail. This can be used only in the vicinity of the star (sun). The velocity of the ship at the start point is zero, but due to the solar wind is $v=4\cdot 10^5\rm{m/s},$ with density of charged protons $\eta_{p}=6\cdot 10^6\rm{m}^{-3}$ (proton density measured by Mariner 2 was $5.4\cdot 10^6\rm{m}^{-3}$\cite{solar}). 
\item
The second example refers to acceleration beyond the Heliosphere region. Density of charged particles in this interstellar region is $\eta_{p}=6\cdot 10^4\rm{m}^{-3}.$ The velocity of the ship (due to interstellar gas) can again be taken as $v=4\cdot 10^5\rm{m/s}.$ Trapped ions will create a force against the engines.
\item
In the third example we will consider deceleration of ship near the target star. The spaceship will use the effect of a space parachute.    
The initial condition of this case will be taken as $\eta_{p}=6\cdot 10^5\rm{m}^{-3},$ $v=4\cdot 10^7\rm{m/s}.$
\end{itemize}
As parameters describing the ship will be taken:
\begin{itemize}
\item
Area of the magnetic field: $S=10^{12}\rm{m}^2$
\item
Mass of the ship $M=10^5\rm{kg}$
\item
Mass of the proton $m_{0}=1.67\cdot 10^{-27}\rm{kg}$
\item
Efficiency of the fusion $0.7\%$ ($m'_{0}=0.993~m_{0}$)
\item
Energy lost $0\%$ ($E_{t}=0$)
\item
Speed of light $c=3\cdot 10^{8}\rm{m/s}$
\end{itemize}

Acceleration of the ship is for the first example $1.4\rm{ms}^{-2}$, second example $0.014\rm{ms}^{-2},$ third example $29.5\rm{ms}^{-2}.$

\medskip
From these values it is obvious that the acceleration is strongly dependent on the actual velocity and density of the protons. A problem with the space umbrella is that, at large distances from stars, there is a very low density of protons. This problem can be partially solved by ionization of the neutral hydrogen atoms. The density of interstellar hydrogen atoms \linebreak $\eta_{H}=(1-2)\cdot 10^5\rm{m}^{-3}$ \cite{Izmodenov:1998ws} is higher than the density of protons. Ionization can be done by a strong magnetic field \cite{0953-4075-44-18-184003} or by a thin 
foil placed in front of magnetic umbrella \cite{whit}. Is it a question if, in the future, man will be able to make such a large and intensive magnetic field for space travel.
\bigskip

\begin{figure}[h]
  \centering
  \begin{minipage}[c]{0.33\textwidth}
    \centering
    \includegraphics[width=4cm, height=4cm, trim=2cm 7cm 6cm 6cm]{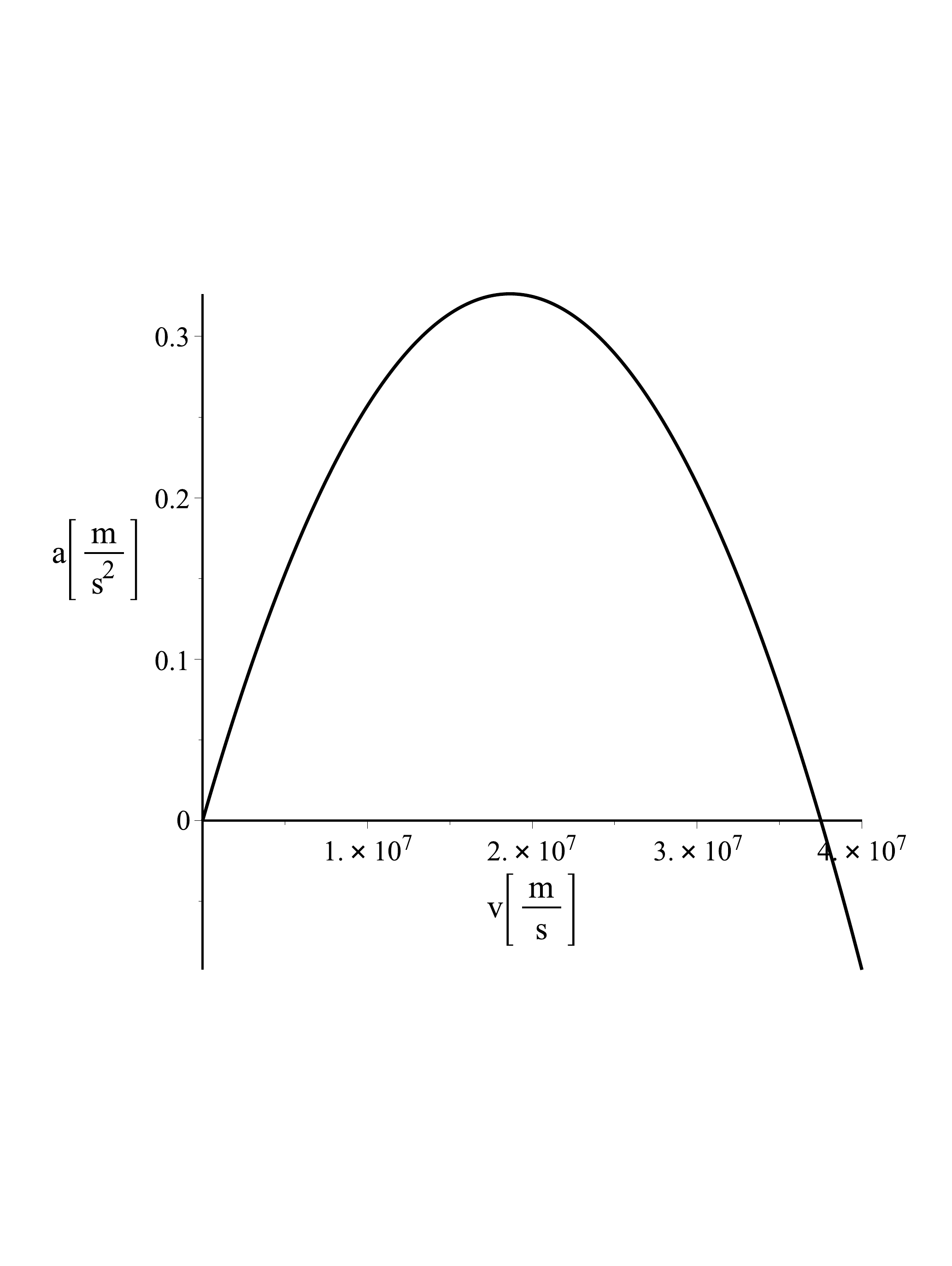}
  \end{minipage}
  \hspace{3cm}
  \begin{minipage}[c]{0.33\textwidth}
    \centering
    \includegraphics[width=4cm, height=4cm, trim=6cm 6cm 2cm 6cm]{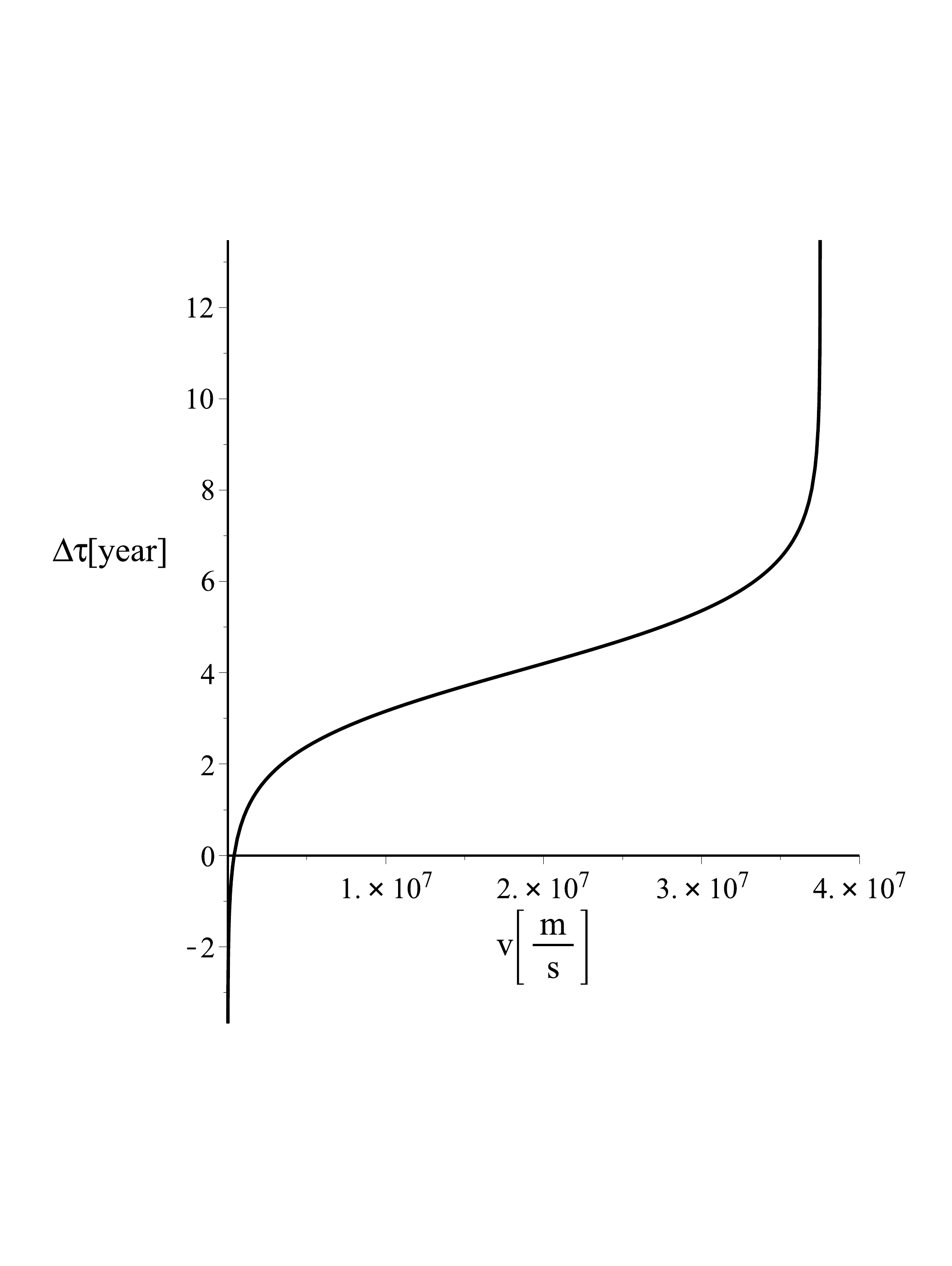}
    \end{minipage}
\\
  \begin{minipage}[c]{0.4\textwidth}
    \centering
    \caption{Acceleration ($a$) as function of velocity ($v$).}
    \label{obrazek1}
  \end{minipage}
  \hspace{2cm}
  \begin{minipage}[c]{0.4\textwidth}
    \centering
    \caption{Time of acceleration ($\Delta\tau$) as function of velocity ($v$).}
    \label{obrazek2}
  \end{minipage}
\end{figure}

Acceleration as a function of velocity is plotted in Figure \ref{obrazek1} (see equation (\ref{6})). Flight duration ($\Delta\tau=\tau_{v_{fin}}-\tau_{v_{0}}$) as a function of velocity ($v_{fin}$) is plotted in Figure \ref{obrazek2} (see equation (\ref{6x})). Initial conditions are: Density of the protons $\eta_{p}=6\cdot 10^4\rm{m}^{-3}$ and initial velocity $v_{0}=4\cdot 10^5\rm{m/s}$ (only for figure \ref{obrazek2}).

From fig.~\ref{obrazek1} we can see that maximal acceleration will be around $0.3\rm{m/s}^{2}$ for $v=6\%c$ and maximal velocity is around $13\%c.$ The ship can be accelerated close to maximal velocity after $7$ years of flight (see fig.~\ref{obrazek2}).

With this maximal velocity it can take approximately $35$ years to travel to the Centauri system. Acceleration (to the speed of the solar wind) near the sun (because of the high density of protons and helpful force from the solar wind) will take only several weeks. The whole voyage can take $40-50$ years.

\section{Standard fusion drive}
In this section we will calculate the mass of hydrogen fuel required to achieve maximal velocity $V_{max}=0.125c$ and then decelerate to zero.
The equation describing the required mass of fuel of a relativistic rocket was found in \cite{rr}
\begin{equation}
M=M_{0}\left(\frac{1+\frac{V_{max}}{c}}{1-\frac{V_{max}}{c}}\right)^{-\frac{c}{2v'}},\label{rak}
\end{equation} 
where $M_{0}$ is the initial mass (at the time when $V=0$) of the rocket and $M$ is the finite mass (at the time when $V=V_{max}$). As we derived before (from equations (\ref{v}) ) velocity of the particles leaving rocket engines is 
\begin{equation}
v'\approx\frac{MV_{2}c}{\sqrt{(m'_{0}c)^{2}+(MV_{2})^{2}}},\label{va}
\end{equation}
where velocity $V_{2}$ is given by (\ref{4}). Combination equations (\ref{4}) and (\ref{va}) gives
\begin{equation}
v'\approx\frac{\sqrt{\left(m_{0}c^{2}-E_{t}\right)^{2}-\left( m'_{0}c^{2}\right)^{2}}}{\left(m_{0}c-\frac{E_{t}}{c}\right)}.\label{vak}
\end{equation}

For $E_{t}=0$ we can use the equation (\ref{vak}) written in the simple form
$$
v'\approx c\sqrt{1-\left( \frac{m'_{0}}{m_{0}}\right)^{2}}.
$$

As was said before, the efficiency of the fusion reaction is $\frac{m'_{0}}{m_{0}}=0.993.$
From the above values the proportion of weight necessary to reach $0.125c$ is equal to $M_{0}/M=2.90.$ In this case the proportion of weight needed for acceleration and deceleration together is $M'_{0}/M=(M_{0}/M)^2=8.40.$ 
This proportion is relatively small in comparison with the proportion of rocket fuel$/$payload of conventional spacecraft needed to reach low earth orbit.

\section{Conclusion} 
In this article we have studied the applicability of a magnetic umbrella in interstellar transport. Magnetic umbrellas work as a trap for charged particles. In the nuclear fusion reactor nuclear energy is utilized to speed up the ship. We expressed equations describing maximal velocity and acceleration of the spaceship and time period of the acceleration. From our results it is obvious that the effect of our system is strongly dependent on the density of protons and the actual velocity of incoming particles. The density is much higher in the vicinity of a star than in interstellar regions. On the other hand, maximal velocity is independent of density of protons, but strongly dependent on the efficiency or on the manner of obtaining the energy.

We calculated concrete examples of using a interstellar ramjet. From our calculations it is obvious that the efficiency of the process is very good and usable for the initial acceleration and finite deceleration near a star. Calculations show that the use of a interstellar ramjet is effective if its weight is comparable with mass of the payload and engines. In the case of the weight of the interstellar ramjet exceeding $8-9$ times the weight of the payload, greater efficiency is obtained through the use of a standard fuel tank with hydrogen.

In this article all calculations with numbers are only rough estimations. For more exact calculations we need to know the exact values of density of the solar wind and density of the protons as a function of distance from the sun and target star.

Despite this inaccuracy we proved that a space journey between two stars (sun and Centauri system) using a interstellar ramjet would take a time period of $40-50$ years and is realizable for a highly developed civilization.

\section{Acknowledgments}
Financial support of author by Department of Physical Electronics, Faculty of Science, Masaryk University is acknowledged.


\bibliographystyle{unsrt}
\bibliography{citace}

\end{document}